%% file: main.tex
\documentclass[conference]{IEEEtran}
\IEEEoverridecommandlockouts
\usepackage{cite}
\usepackage{amsmath,amssymb,amsfonts}
\usepackage{algorithmic}
\usepackage{graphicx}
\usepackage{textcomp}
\usepackage{color,soul}
\usepackage{enumerate}
\usepackage{pgfplots}
\usepackage{optidef}
\usepackage{dblfloatfix}
\def\BibTeX{{\rm B\kern-.05em{\sc i\kern-.025em b}\kern-.08em
		T\kern-.1667em\lower.7ex\hbox{E}\kern-.125emX}}
\usepackage{tikz}
\begin{document}
	\input{sections/title}
	\input{sections/abstract}

	\input{sections/introduction}
	\input{sections/VUdefinitions}
	\input{sections/TPOPF}

    \input{sections/Simulation}
	\input{sections/conclusion}
	\bibliography{IEEEfull,references}
\end{document}

%% file: sections/title.tex
\title{On the Impact of Different Voltage Unbalance Metrics in Distribution System Optimization}

\author{
\IEEEauthorblockN{Kshitij Girigoudar, Line A. Roald}
\IEEEauthorblockA{Electrical and Computer Engineering Department,
University of Wisconsin--Madison, Madison, USA\\
\{girigoudar, roald\}@wisc.edu}
\thanks{This work was funded by the U.S. Department of Energy’s Office of Energy Efficiency and Renewable Energy (EERE) under Solar Energy Technologies Office (SETO) Agreement Number 34235.}

}
\maketitle

%% file: sections/abstract.tex
\begin{abstract}
With increasing penetrations of single-phase, rooftop solar PV installations, the relative variations in per-phase loading and associated voltage unbalance are expected to increase. High voltage unbalance may increase network losses and lead to failure of three-phase equipment such as motor loads. However, solar PV panels are connected to the grid through inverters, which can provide reactive power support and may mitigate some of these negative effects. In this paper, we utilize a three-phase AC optimal power flow (OPF) formulation to minimize voltage unbalance using reactive power from solar PV inverters. When considering actions to reduce voltage unbalance, it is important to recognize that various organizations such as IEC, NEMA and IEEE provide different and partially inconsistent definitions of voltage unbalance in their power quality standards. This paper analyzes the impact of the different voltage unbalance metrics using different combinations of voltage unbalance objectives and constraints. For our analysis, the optimization scheme is tested on two unbalanced low-voltage distribution networks. We observe that minimizing voltage unbalance defined by one standard might actually increase voltage unbalance as defined by another standard, potentially resulting in equipment damage. We also observe that minimizing voltage unbalance does not always lead to lower network losses. However, considerable reduction in voltage unbalance with low network losses can be achieved by minimizing the losses while simultaneously enforcing limits on multiple definitions of voltage unbalance. 
\end{abstract}


%% file: sections/introduction.tex
\section{Introduction}
\label{sec:Introduction}
Distributed energy resources (DER) have been gaining popularity in the past few years. With increasing penetrations of highly variable solar PV generation, which are typically not allocated equally between phases~\cite{karimi2016photovoltaic}, the relative variations in per-phase loading and associated voltage unbalance are expected to increase. While three-phase equipment such as three-phase motor loads are designed to withstand moderate levels of voltage unbalance, significant unbalance may lead to costly damage or derating of motors in order to avoid premature failure \cite{muljadi1985induction}. Additionally, voltage unbalance also leads to higher network losses, which increases cost and may damage utility assets such as transformers \cite{woolley2012statistical}.

Various organizations such as IEC~\cite{standard2002}, NEMA~\cite{ANSINEMAStandard.1993}, and IEEE~\cite{398556} define voltage unbalance in their power quality standards. They also describe the maximum limits of voltage unbalance for both grid operators, who  need to maintain high power quality throughout the network, as well as for equipment manufacturers, who design equipment to withstand voltage unbalance without any persistent damage. However, the definitions are different and partially inconsistent, and voltage unbalance levels considered acceptable by one standard may violate limits defined by another standard \cite{pillay2001definitions}. 

Conventional methods to address voltage unbalance problems include changing residential loads or DER connections between phases, which is typically done at most few times a year \cite{shahnia2014voltage}, or installation of distribution static synchronous compensators \cite{xu2010voltage}, which might be expensive. Other methods include utilizing transformer online tap changer (OLTC), voltage regulators and switched capacitors. However, frequent switching of these devices can cause failures in the distribution system \cite{liu2012coordinated}. More recently, inverter-based reactive power control has gained popularity due to better transient performance and lower additional investments compared to the conventional methods \cite{su2014optimal}. Solar PV inverters typically do not operate at maximum power rating, and the additional available capacity can be used to absorb or inject reactive power to the grid. There are a number of papers in the existing literature that propose mitigation of voltage unbalance in distribution systems with PV inverters. In \cite{weckx2014reducing}, an optimization problem is solved using only active power of inverters as control variables and reactive power is ignored. Controllers based on three-phase optimal power flow formulation that include reactive power support are proposed in \cite{su2014optimal}, \cite{karagiannopoulos2018centralised} and \cite{schneider2017analytic}, but they implement only the IEC voltage unbalance definition either as objective or constraints in the problem formulation. The IEEE definition is used in \cite{bajo2015voltage} to minimize unbalance in distribution systems with high penetration of solar PV installations.

While previous literature has looked at methods to mitigate voltage unbalance by choosing one of many metrics, this paper takes a more comprehensive approach and investigates how the \emph{definition of voltage unbalance} impact the solution of the optimization problem. 
We show that minimizing voltage unbalance with respect to one definition may increase the voltage unbalance with respect to another definition, and also have different impacts on other important system parameters such as network losses. 
Using a detailed three-phase distribution system model, we solve a three-phase AC optimal power flow (TP-OPF) problem which minimizes voltage unbalance by utilizing the reactive power injections of the inverters. In summary, the major contributions of this paper are: 
\begin{enumerate} 
\item The formulation of a TP-OPF problem which is flexible enough to represent the different voltage unbalance definitions either as constraints or objective function, and also include models of major system components such as transformers and loads in any configuration (wye or delta). 
\item An in-depth comparison of the results obtained by minimizing different voltage unbalance definitions, which show that minimizing unbalance using one standard might violate the limits set by another standard or lead to increased network losses. 
\end{enumerate}  
    
The rest of the paper is structured as follows: Section~ \ref{sec:VUdef} introduces the commonly used voltage unbalance definitions and other important power quality metrics in distribution systems. Section~\ref{sec:TP_OPF} provides a summary of the modelling of various distribution system components as well as the TP-OPF problem formulation. Section~\ref{sec:Results} describes the simulation results for two test feeders. Section~\ref{sec:Conclusion} concludes the paper.

%% file: sections/VUdefinitions.tex
\section{Power Quality Metrics in Distribution systems}
	\label{sec:VUdef}
	This section introduces the various performance metrics that we use later to evaluate the power supply quality of distribution systems. This includes the voltage unbalance definitions as well as other important metrics such as network losses and power factor at the substation that a distribution system operator (DSO) must manage to operate the network efficiently. 
    
    \subsection{Voltage Unbalance Definitions}	
	Voltage unbalance is a condition in the three-phase power system  when either the phase voltages have asymmetric magnitudes, the phase angle displacement is not equal to $120^ \circ$, or a combination of both \cite{kim2005comparison}. In this subsection, we summarize the three commonly used definitions for voltage unbalance from IEC~\cite{standard2002}, NEMA~\cite{ANSINEMAStandard.1993}, and IEEE~\cite{398556} along with some interesting observations from \cite{VUdefcompare}. 
	
    The following notation is used to represent the voltage phasors: $\mathbf{V} = V\angle{\theta}$, where $\mathbf{V}$ denotes the phasor, $V$ represents the magnitude, and $\angle{\theta}$ is the phase angle. The complex conjugate of $\mathbf{V}$ is denoted by $\overline{\mathbf{V}}$. Let $j = \sqrt{-1}$. The element-wise product is represented using $\odot$. 
    
	\subsubsection{IEC definition (VUF)}
	 Voltage Unbalance Factor (VUF) is described using symmetrical components, or more specifically defined based on the  magnitudes of the negative and positive sequence voltages. The VUF definition is adopted by the IEC standard \mbox{61000-2-2}~\cite{standard2002} and also referred to as the ``true'' definition. It is given by
	\begin{align}
	    & VUF ~ [\%] = \frac{|\mathbf{V}_n|}{|\mathbf{V}_p|} \times 100,   \label{eq:VUF} \quad\mathrm{where} \\[3pt]
	    & \mathbf{V}_p \!=\! \frac{\mathbf{V}_a+\textbf{a}\!\cdot\!\mathbf{V}_b+\textbf{a}^2\!\cdot\!\mathbf{V}_c}{3},~ 
	     \mathbf{V}_n \!=\! \frac{\mathbf{V}_a+\textbf{a}^2\!\cdot\!\mathbf{V}_b+\textbf{a}\!\cdot\!\mathbf{V}_c}{3}. \notag 
	\end{align} 
	Here, $\mathbf{V}_n$ and $\mathbf{V}_p$ are the negative and positive sequence voltage phasors, respectively; $\textbf{a}=1\angle120^\circ$; and $\mathbf{V}_a$, $\mathbf{V}_b$, $\mathbf{V}_c$ are the three-phase line-to-ground voltage phasors. The IEC standard~\cite{standard2002} requires that the voltage unbalance defined in~\eqref{eq:VUF} should be less than $2$\% for low- and medium-voltage systems.

	\subsubsection{NEMA definition (LVUR)}
	 The NEMA definition of voltage unbalance~\cite{ANSINEMAStandard.1993}, which is also referred to as the Line Voltage Unbalance Rate (LVUR), is used by motor manufacturers. It calculates unbalance using the line-to-line voltage magnitudes $V_{ab}, V_{bc}$ and $V_{ca}$:
	\begin{align}
	    \label{eq:LVUR}
	    & \hspace*{50pt} LVUR \;[\%] = \frac{\Delta V_{L}^{max}}{V_{L}^{avg}} \times 100,\\
	    \nonumber & \hspace*{50pt} \mathrm{where}\;\; V_{L}^{avg} = \frac{V_{ab}+V_{bc}+V_{ca}}{3}, \\
	    \nonumber & \Delta V_{L}^{max} = \max\{|V_{ab}-V_{L}^{avg}|,|V_{bc}-V_{L}^{avg}|,|V_{ca}-V_{L}^{avg}|\}. 
		\end{align} 
	To comply with the NEMA MG-1 \cite{ANSINEMAStandard.1993} and ANSI C84.I \cite{ansi1995} standards, the maximum voltage unbalance, as defined in~\eqref{eq:LVUR}, must not exceed $3$\% under no-load conditions. For voltage unbalance greater than 1\%, the induction motors should be derated by an appropriate factor \cite{ANSINEMAStandard.1993}. 

	\subsubsection{IEEE definition (PVUR)}
	The IEEE definition in~\cite{398556}, which is commonly referred to as Phase Voltage Unbalance Rate (PVUR), is calculated using the line-to-ground voltage magnitudes $V_{a}, V_{b}$ and $V_{c}$:
	\begin{align}
	\label{eq:PVUR}
	& \hspace*{50pt} PVUR\; [\%] = \frac{\Delta V_{P}^{max}}{V_{P}^{avg}} \times 100,\\
	\nonumber & \hspace*{50pt} \mathrm{where}\;\;V_{P}^{avg} = \frac{V_{a}+V_{b}+V_{c}}{3},\\
	& \nonumber \Delta V_{P}^{max} = \max\{|V_{a}-V_{P}^{avg}|,|V_{b}-V_{P}^{avg}|,|V_{c}-V_{P}^{avg}|\}.
	\end{align} 
	The IEEE standard 141-1993~\cite{398556} prescribes that phase-voltage unbalances should be limited to be below $2$\% to avoid overheating of motors near full load conditions. 
	
	Several previous works have investigated how the different voltage unbalance metrics \eqref{eq:VUF}-\eqref{eq:PVUR} compare. In particular, \cite{VUdefcompare} aimed at deriving analytical relationships that allow us to bound the value of VUF using measurements of LVUR and PVUR. For reasonable levels of voltage unbalance, it was shown that VUF is bounded by LVUR using the following relationship:
	\begin{align}
	    LVUR \lesssim VUF \lesssim \left(\frac{2}{\sqrt{3}}\right)\cdot LVUR.
	\end{align}
	It was also observed that PVUR does not have any direct relationship with either LVUR or VUF because it is computed based on the voltage magnitude of each phase and does not include any information about the angle between phases. In comparison, LVUR includes an approximate measure of the phase angle unbalance since it relies on line-to-line voltage magnitude measurements.
    
    Different practical considerations are important for the definition of the voltage unbalance metrics. It can be challenging to measure VUF as it requires access to both the voltage magnitudes and relative phase angles. 
	In a distribution network with limited availability of voltage angle measurements, it is therefore more practical to measure unbalance using either LVUR or PVUR, since these definitions only require voltage magnitude measurements that can be obtained using standard RMS meters. 
	
    \subsection{Network Losses}
    Operating the network efficiently with minimal losses is an important objective for distribution system operators, as the losses is a main driver for the cost of electricity distribution.    The network losses can be calculated as
    \begin{align}
            P_{loss} = \sum_{l=1}^{n_{br}} \Re\{\mathbf{V}_l \cdot \overline{\mathbf{I}}_l\}, \label{eq:loss}
    \end{align}     
    where $\mathbf{V}_l$ is the voltage drop across branch $l$ and $\mathbf{I}_l$ is the current through branch $l$ for a network with $n_{br}$ branches.
    
    \subsection{Power Factor}
    Another important objective for the distribution system operator is to maintain a high power factor at the point of interconnection to the transmission system, which avoids incurring additional cost of procuring the higher apparent power \cite{deaver2010method} from transmission system operators. A low power factor at the substation might also require additional infrastructure to distribute the additional apparent power \cite{deaver2010method}, thereby increasing the investment cost. Furthermore, lower power factor leads to higher currents flowing in the distribution network which increases losses.  
    We measure the power factor at the substation as the ratio of real to apparent power given by
        \begin{align}
            cos\,\phi_{ss} = \frac{P_G^a+ P_G^b+P_G^c}{S_G^a+ S_G^b+S_G^c}, \label{eq:pf}
    \end{align} 
     where $P_G^a,~P_G^b,~P_G^c$ and $S_G^a,~S_G^b,~S_G^c$ are the per-phase active and apparent power injections at the distribution substation, respectively.
    

%% file: sections/TPOPF.tex
\section{Three-Phase Optimal Power Flow}
	\label{sec:TP_OPF}
	 In this section, we describe the formulation for a three-phase optimal power-flow (TP-OPF) problem, which was developed in the polar coordinate system. We consider a network with $n_b$ buses and $n_{br}$  branches.  Additionally, there are  $n_g$ generators as well as $n_{l}$ loads in the system. The vector notation used to define variables in the TP-OPF formulation along with other parameters for the $i^{th}$ bus with phase $a,b,c$ is
\begin{align}
     &V_{i}^{abc} = {\begin{bmatrix}
                        V_{i}^{a} & V_{i}^{b} & V_{i}^{c}
                    \end{bmatrix}}^\intercal, ~
     P_{G,i}^{abc} = {\begin{bmatrix}
                        P_{G,i}^a & P_{G,i}^b & P_{G,i}^c
                    \end{bmatrix}}^\intercal, \\               
     &\theta_{i}^{abc} = {\begin{bmatrix}
                        \theta_{i}^{a} & \theta_{i}^{b} & \theta_{i}^{c}
                    \end{bmatrix}}^\intercal, ~
     Q_{G,i}^{abc} = {\begin{bmatrix}
                        Q_{G,i}^a & Q_{G,i}^b & Q_{G,i}^c
                    \end{bmatrix}}^\intercal,   \notag             
\end{align}
where $V_{i}^{abc}$ and $\theta_{i}^{abc}$ are the line-to-ground voltage magnitude and angle variables, respectively, at every bus $i=1,2,\hdots n_b$. The active and reactive power variables at every generator bus $i=1,2,\hdots n_g$ are denoted by $ P_{G,i}^{abc}$ and $ Q_{G,i}^{abc}$, respectively. 

Typically, the distribution systems have considerable number of single and two-phase buses. In such cases, no variable is assigned to the missing phases. For example, if bus $i$ has only two phases without phase $C$, the voltage magnitude variable will be $V_{i}^{abc} = [V_{i}^{a} ~~ V_{i}^{b}]^\intercal$ and the voltage angle variable is $\theta_{i}^{abc}=[\theta_{i}^{a} ~~ \theta_{i}^{b}]^\intercal$.   The following sections describe the mathematical models for distribution system components, objective functions, and constraints. 

\subsection{Distribution Lines and Cables}
The $\pi$-model is used for representation of distribution lines and cables \cite{bazrafshan2018comprehensive}. This model includes the series impedance as well as the shunt admittance of the line or cable. For any three-phase branch between the $i^{th}$ and $j^{th}$ bus, the branch admittance matrix $Y_{ij}^{abc}\in \mathbb{R}^{3\times 3}$ 
is denoted as  
\begin{align}
    Y_{ij}^{abc}   
    &=  {
    \underbrace{\begin{bmatrix}
    z_{ij}^{aa} & z_{ij}^{ab} & z_{ij}^{ac} \\
    z_{ij}^{ba} & z_{ij}^{bb} & z_{ij}^{bc} \\
    z_{ij}^{ca} & z_{ij}^{cb} & z_{ij}^{cc} \\
    \end{bmatrix}}_{Z_{ij,se}^{abc}}}^{-1} + 
    \begin{bmatrix}
    j\cdot b_{ij}^{aa} & 0 & 0 \\
    0 & j\cdot b_{ij}^{bb} & 0 \\
    0 & 0 & j\cdot b_{ij}^{cc} \\
    \end{bmatrix} \notag \\
    &= Y_{ij,se}^{abc} + Y_{ij,sh}^{abc}. \label{eq:Ypr} 
\end{align}
 The diagonal entries of the series impedance matrix $Z_{ij,se}^{abc}$ are the self impedance values in each phase while the off-diagonal entries represent mutual impedance to account for inter-phase coupling \cite{kersting2006distribution}. The shunt capacitive susceptance between the branches is represented by $Y_{ij,sh}^{abc}$. 
 
When the branch $ij$ has less than three phases, the branch admittance matrix $Y_{ij}^{abc}$ takes on a reduced form. As an example, if bus $i$ has three phases $abc$ while bus $j$ only has two phases $ab$, and the branch $ij$ between them also has two phases $ab$, then $Y_{ij}^{abc}\in \mathbb{R}^{3\times 2}$ is defined as 
\begin{align}
    Y_{ij}^{abc}   
    &=  \begin{bmatrix}
    y_{ij}^{aa} & y_{ij}^{ab} \\
    y_{ij}^{ba} & y_{ij}^{bb} \\
    0 & 0 \\
    \end{bmatrix} + 
    \begin{bmatrix}
    j\cdot b_{ij}^{aa} & 0 \\
    0 & j\cdot b_{ij}^{bb} \\
    0 & 0  \\
    \end{bmatrix}. \notag 
\end{align}
In general, the dimension of this matrix is given by $Y_{ij}^{abc}\in \mathbb{R}^{n_{i}\times n_{j}}$, where $n_i$ and $n_j$ are the number of phases at bus $i$ and $j$, respectively. 
 
 The overall system bus admittance matrix $\mathbf{Y}_{bus}$ can be defined using the branch admittance matrices $Y_{ij}^{abc}$, 
\begin{align}
    \mathbf{Y}_{bus} &= G^{abc} + j \cdot B^{abc} \label{eq:Ybus}\\
     &=
      \begin{bmatrix}
        \sum_{k=1}^{n_b}Y_{1k}^{abc} & -Y_{12}^{abc} & \hdots & -Y_{1n_b}^{abc}  \\[0.2cm] 
        -Y_{21}^{abc} & \sum_{k=1}^{n_b}Y_{2k}^{abc} & \hdots & -Y_{2n_b}^{abc} \\
        \vdots & \vdots & \ddots & \vdots \\
        -Y_{n_b1}^{abc} & -Y_{n_b2}^{abc} & \hdots & \sum_{k=1}^{n_b}Y_{n_bk}^{abc}
     \end{bmatrix}. \notag
\end{align}
Here, the matrices $G^{abc}$ and $B^{abc}$ corresponding to the real and imaginary part of $\mathbf{Y}_{bus}$ are the bus conductance and susceptance matrices, respectively. 
In the power flow equations, we will refer to the submatrices of \eqref{eq:Ybus} that describe the admittance between bus $i$ and $j$ as $G_{ij}^{abc}$ and $B_{ij}^{abc}$, respectively.

\subsection{Transformer}
A two-winding, three-phase transformer \cite{arrillaga1983computer} has been modeled with the following assumptions:
\begin{enumerate}[(i)] 
\item Transformer bank consists of three single-phase transformers.
\item The three sets of coils have similar characteristics with the admittance values for primary and secondary windings along with couplings equal to the transformer leakage admittance $y_t$.
\end{enumerate} 

The three-phase transformers are represented using the transformer admittance matrix $Y_{\textbf{T}}^{abc} \in\mathbb{R}^{6\times 6}$   where
\begin{align}
\mathbf{Y}_{T} = 
     \begin{bmatrix}
      Y_{ii}^{abc} & Y_{ij}^{abc}  \\
      Y_{ji}^{abc} & Y_{jj}^{abc}  \\
     \end{bmatrix} ~\forall \, i,j \in \{1,2,\hdots n_b\}.
     \label{eq:TFmodel}
\end{align}
The four sub-matrices of $\mathbf{Y}_{T}$ can be calculated similar to the branch admittance matrix in~\eqref{eq:Ypr}. Table \ref{TF_conn} illustrates the sub-matrices of $\mathbf{Y}_{T}$ for various IEC transformer
connections \cite{bazrafshan2018comprehensive} where
\begin{align}
&Y_{I} = \begin{bmatrix}
        1 & 0 & 0 \\
        0 & 1 & 0 \\
        0 & 0 & 1 \\
        \end{bmatrix} \cdot y_t, \;\;
Y_{II} = 
        \begin{bmatrix}
        2 & -1 & -1 \\
        -1 & 2 & -1 \\
        -1 & -1 & 2 \\
        \end{bmatrix} \cdot \frac{y_t}{3} \notag\\
&Y_{III} = 
        \begin{bmatrix}
        -1 & 1 & 0 \\
        0 & -1 & 1 \\
        1 & 0 & -1 \\
        \end{bmatrix}\cdot \frac{y_t}{\sqrt{3}}.
\end{align}

\begin{table}[h]
\renewcommand{\arraystretch}{1.4}
\centering
\caption{Sub-matrices for transformer connections}
\begin{tabular} {|c|c|c|c|c|c|c|}
\hline
{} & \textbf{YNyn0} & \textbf{Yy0} & \textbf{YNd1} & \textbf{Yd1} & \textbf{Dyn1} &  \textbf{Dyn11}   \\ \hline
Y$_{ii}^{abc}$ & Y$_{I}$ & Y$_{II}$ & Y$_{I}$ & Y$_{II}$ & Y$_{II}$ & Y$_{II}$ \\ \hline
Y$_{jj}^{abc}$ & Y$_{I}$ & Y$_{II}$ & Y$_{II}$ & Y$_{II}$ & Y$_{I}$ & Y$_{I}$ \\ \hline
Y$_{ij}^{abc}$ & -Y$_{I}$ & -Y$_{II}$ & Y$_{III}$ & Y$_{III}$ & Y$_{III}$ & Y$^\intercal_{III}$ \\ \hline
Y$_{ji}^{abc}$ & -Y$_{I}$ & -Y$_{II}$ & Y$^\intercal_{III}$ & Y$^\intercal_{III}$ & Y$^\intercal_{III}$ & Y$_{III}$ \\ \hline
\end{tabular}
\label{TF_conn}
\end{table} 

\subsection{Voltage Regulator}
Regulators are considered as a special case of the transformer model with YNyn0 connection type and \emph{constant} tap ratios defined by $t_a,\, t_b$ and $t_c$ for each phase \cite{wang2017accurate}. The regulator admittance matrix is calculated similar to the transformer admittance matrix in~\eqref{eq:TFmodel} with the following modifications:
\begin{enumerate}[(i)]
    \item Multiply the three diagonal entries of the primary winding self-admittance (Y$_{ii}^{abc}$) by $t_a^2,\, t_b^2$ and $t_c^2$, respectively
    \item Multiply the three diagonal entries of the mutual admittance (Y$_{ij}^{abc}$, Y$_{ji}^{abc}$) by $-t_a,\, -t_b$ and $-t_c$, respectively
\end{enumerate}

\subsection{Distribution Substation}
The distribution substation is chosen as a slack bus to provide a reference for the measurement of voltage angles. For the slack bus $k$, we enforce voltage constraints as 
\begin{align}
        V_{k}^{abc}\angle \theta_{k}^{abc} = 
    {\begin{bmatrix}
        {1\angle 0^\circ} & 
        {1\angle-120^\circ} & 
        {1\angle 120^\circ}
    \end{bmatrix}}^\intercal.  \label{eq:Vref_constr}
\end{align}
For any other bus $i \in \{1,2,\hdots n_b\}\backslash k$, the voltage magnitude limits are 
\begin{align}
    V_{i,min}^{abc} \leq V_{i}^{abc} \leq V_{i,max}^{abc}. \label{eq:V_constr}
\end{align}
The distribution substation also serves as a large-scale generator with power injection limits given by
\begin{subequations}
\begin{align}
    {P_{G,k}^{abc}}_{min} \leq P_{G,k}^{abc} \leq {P_{G,k}^{abc}}_{max},\\
    {Q_{G,k}^{abc}}_{min} \leq Q_{G,k}^{abc} \leq {Q_{G,k}^{abc}}_{max}.
\end{align} \label{eq:PQref_constr}
\end{subequations}

\subsection{Solar PV Inverters}
Our main source of flexibility in the distribution grid are the single-phase, solar PV inverters which are modelled as power injection sources at buses $j \in \{1,2,\hdots n_g\}$. The controllable variables of these single-phase PV inverters connected to phase $\phi \in \{a,b,c\}$ at bus $j$ is the reactive power injection $Q_{G,j}^\phi$ which can be constrained \cite{su2014optimal} as 
\begin{align}
   -\sqrt{{S_{G,j}^\phi}^2-{P_{G,j}^\phi}^2} \leq Q_{G,j}^\phi \leq \sqrt{{S_{G,j}^\phi}^2-{P_{G,j}^\phi}^2}, \label{eq:inv_pow}
\end{align}
where $S_{G,j}^\phi$ and $P_{G,j}^\phi$ are the maximum apparent power limit and specified active power of the inverter connected to phase $\phi$ at bus $j$, respectively. It is important to note that we use the available capacity for reactive power control only. So, we do not control or curtail the active power of the inverter.

\subsection{ZIP Loads}
A polynomial load (ZIP) model is used to represent various loads as a function of the voltage magnitude $V$ \cite{bazrafshan2018comprehensive}. The complex power, $S_{D}$ is defined as 
\begin{align}
S_{D} = P_{D}+j\cdot Q_{D} = \Big(& P_P+P_I\cdot \frac{V}{V_{b}}+P_Z\cdot {\frac{V^2}{V_{b}^2}}\Big) \notag\\
     + j\cdot \Big(& Q_P+Q_I\cdot \frac{V}{V_{b}}+Q_Z\cdot {\frac{V^2}{V_{b}^2}}\Big),  \label{eq:load}
\end{align}
where $P_P+ j Q_P $ is the constant power load component, $P_I+ j Q_I $ is the constant current load component at base voltage, $V_{b}$ and $P_Z+ j Q_Z $ is the constant impedance load component at base voltage, $V_{b}$. Note that bus shunts are included in our model as constant impedance loads.

\subsection{Nodal Power Balance Constraints}
Conservation of power must be satisfied at every bus $i=1,2,\hdots n_b$ in the system. The nodal power balance constraints are expressed as
\begin{subequations}
\begin{align}
    &P_i^{abc} = P_{G,i}^{abc}-P_{D,i}^{abc},\\
    &= V_i^{abc}\odot\sum_{j=1}^{n_b}\Big[G_{ij}^{abc}\odot C\Big(\theta_{ij}^{abc}\Big)+B_{ij}^{abc}\odot S\Big(\theta_{ij}^{abc}\Big) \Big] \cdot V_j^{abc}, \notag\\
    &Q_i^{abc} = Q_{G,i}^{abc}-Q_{D,i}^{abc},\\
    &= V_i^{abc}\odot\sum_{j=1}^{n_b}\Big[G_{ij}^{abc}\odot S\Big(\theta_{ij}^{abc}\Big)-B_{ij}^{abc}\odot C\Big(\theta_{ij}^{abc}\Big) \Big]\cdot V_j^{abc}, \notag 
\end{align} \label{eq:pow_bal}
\end{subequations}
where $P_{G,i}^{abc}$, $Q_{G,i}^{abc}$ represent the real and reactive components of the substation and solar PV inverters at bus $i$, respectively, and $P_{D,i}^{abc}$, $Q_{D,i}^{abc}$ denote the real and reactive components of the load demand, respectively. $G_{ij}^{abc}$ and $B_{ij}^{abc}$ are the conductance and susceptance sub-matrices, respectively, as defined in \eqref{eq:Ybus}. $C\Big(\theta_{ij}^{abc}\Big)$ and $S\Big(\theta_{ij}^{abc}\Big)$ are defined as
\begin{align}
C\Big(\theta_{ij}^{abc}\Big)= 
  \begin{bmatrix}
    cos(\theta_{i}^{a}-\theta_{j}^{a}) & cos(\theta_{i}^{a}-\theta_{j}^{b}) & cos(\theta_{i}^{a}-\theta_{j}^{c}) \\
     cos(\theta_{i}^{b}-\theta_{j}^{a}) & cos(\theta_{i}^{b}-\theta_{j}^{b}) & cos(\theta_{i}^{b}-\theta_{j}^{c}) \\
     cos(\theta_{i}^{c}-\theta_{j}^{a}) & cos(\theta_{i}^{c}-\theta_{j}^{b}) & cos(\theta_{i}^{c}-\theta_{j}^{c}) 
    \end{bmatrix},  \notag
\end{align} 
\begin{align}
S\Big(\theta_{ij}^{abc}\Big)= 
  \begin{bmatrix}
    sin(\theta_{i}^{a}-\theta_{j}^{a}) & sin(\theta_{i}^{a}-\theta_{j}^{b}) & sin(\theta_{i}^{a}-\theta_{j}^{c}) \\
     sin(\theta_{i}^{b}-\theta_{j}^{a}) & sin(\theta_{i}^{b}-\theta_{j}^{b}) & sin(\theta_{i}^{b}-\theta_{j}^{c}) \\
     sin(\theta_{i}^{c}-\theta_{j}^{a}) & sin(\theta_{i}^{c}-\theta_{j}^{b}) & sin(\theta_{i}^{c}-\theta_{j}^{c}) 
    \end{bmatrix}.  \notag
\end{align} 

\subsection{Voltage Unbalance Minimization}
This subsection describes the formulation of different voltage unbalance metrics as constraints or objective function. Note that our formulation only defines voltage unbalance at three-phase buses.

\subsubsection{Voltage Unbalance Factor (VUF)}
We express VUF as a function of our optimization variables by defining 
\begin{align}
    VUF_i = \frac{|\mathbf{V}_{n,i}|}{|\mathbf{V}_{p,i}|} = \sqrt{\frac{{V_{n,i}^{d}}^2+{V_{n,i}^{q}}^2}{{V_{p,i}^{d}}^2+{V_{p,i}^{q}}^2}} 
    \label{eq:VUFexpr}
\end{align} 
where $\mathbf{V}_{n,i}, \mathbf{V}_{p,i}$ are defined according to~\eqref{eq:VUF}, and have real and imaginary components defined as $\mathbf{V}_{n,i} = V_{n,i}^{d} + jV_{n,i}^{q}$ and $\mathbf{V}_{p,i} = V_{p,i}^{d} + jV_{p,i}^{q}$, respectively.

We minimize VUF by minimizing the sum of the squared VUF components per three-phase bus, i.e., 
\begin{align}
    \min \sum_{i=1}^{n_b} \frac{{V_{n,i}^{d}}^2+{V_{n,i}^{q}}^2}{{V_{p,i}^{d}}^2+{V_{p,i}^{q}}^2}. \label{eq:VUF_obj}
\end{align}
We can enforce constraints on the VUF at each three-phase bus $i$ by imposing an 
upper bound $u_{VUF}$ on~\eqref{eq:VUFexpr}. By squaring both sides and rearranging terms, we get the following constraint
\begin{align}
    {V_{n,i}^{d}}^2+{V_{n,i}^{q}}^2  \leq u_{VUF}^2 \cdot ({V_{p,i}^{d}}^2+{V_{p,i}^{q}}^2). \label{eq:VUF_const}
\end{align} 

\subsubsection{Phase Voltage Unbalance Rate (PVUR)}
To express PVUR in terms of our optimization variables, we first express the absolute value of voltage deviation. This is done by defining additional variables ${z_{1,i}^{P}}^{abc}$ at each three-phase bus, which represent the absolute voltage magnitude deviation in each phase. This gives rise to the following set of constraints,
\begin{subequations}
\begin{align}
    {z_{1,i}^{P}}^{a}  &\geq V_{a,i} - V_{P,i}^{avg},~ {z_{1,i}^{P}}^{a}  \geq -(V_{a,i} - V_{P,i}^{avg}),\\
    {z_{1,i}^{P}}^{b}  &\geq V_{b,i} - V_{P,i}^{avg}, ~ {z_{1,i}^{P}}^{b}  \geq -(V_{b,i} - V_{P,i}^{avg}),\\
     {z_{1,i}^{P}}^{c}  &\geq V_{c,i} - V_{P,i}^{avg},~ {z_{1,i}^{P}}^{c}  \geq -(V_{c,i} - V_{P,i}^{avg}).
\end{align} \label{eq:PVUR_z1}
\end{subequations}
Next, we define a variable $z_{2,i}^P$ for every bus in the system. It defines the maximum relative voltage deviation across the three phases,
\begin{align}
     z_{2,i}^P \geq \frac{{z_{1,i}^{P}}^{a}}{V_{P,i}^{avg}},~ z_{2,i}^P  \geq \frac{{z_{1,i}^{P}}^{b}}{V_{P,i}^{avg}} ~~ \mathrm{and} ~~ z_{2,i}^P \geq \frac{{z_{1,i}^{P}}^{c}}{V_{P,i}^{avg}}. \label{eq:PVUR_z2}
\end{align}
where $V_{P,i}^{avg}$ is the average of the three phase voltage magnitudes as defined in~\eqref{eq:PVUR}. For buses that are not three-phase buses, we set $z_{2,i}^P=0$.
When minimizing PVUR, we minimize the sum of $z_2^P$ across all buses, i.e.,
\begin{align}
   \min \sum_{i=1}^{n_b} z_{2,i}^P.  \label{eq:PVUR_obj}
\end{align}
To enforce constraints on PVUR, we use
\begin{align}
   z_{2,i}^P \leq u_{PVUR} \label{eq:PVUR_const}
\end{align}
where $u_{PVUR}$ is the specified PVUR limit. 

\subsubsection{Line Voltage Unbalance Rate (LVUR)}
We use a similar approach as for PVUR and define auxiliary variables $z_{1,i}^{L,abc}$ and $z_{2,i}^L$ for all three-phase buses. This allows us to express LVUR on bus $i$ using following set of constraints,
\begin{subequations}
\begin{align}
     {z_{1,i}^{L a}}  &\geq V^{ab}_i - V_{L,i}^{avg},~ {z_{1,i}^{La}} \geq -(V^{ab}_i - V_{L,i}^{avg}),\\
    {z_{1,i}^{Lb}}  &\geq V_i^{bc} - V_{L,i}^{avg}, ~ {z_{1,i}^{Lb}}  \geq -(V_i^{bc} - V_L^{avg}),\\
     {z_{1,i}^{Lc}} &\geq V_i^{ca} - V_{L,i}^{avg},~ {z_{1,i}^{Lc}}  \geq -(V_i^{ca} - V_{L,i}^{avg}),\\
     z_{2,i}^L &\geq \frac{{z_1^{La}}}{V_{L,i}^{avg}},~ z_{2,i}^L  \geq \frac{{z_{1,i}^{Lb}}}{V_{L,i}^{avg}} ~~ \mathrm{and}~~ z_{2,i}^L \geq \frac{{z_{1,i}^{Lc}}}{V_{L,i}^{avg}}. 
\end{align} \label{eq:LVUR_def1}
\end{subequations}
Here, the line-to-line voltage magnitude between phase $a$ and $b$ on bus $i$, denoted by $V_{i}^{ab}$, is calculated from the other optimization variables given by
\begin{align}
    {V_{i}^{ab}}^2 &= V_{i}^{{a}^2}+{V_i^b}^2-2  V^a_i  V^b_i \cdot cos\Big(\theta_i^{a}-\theta_i^{b}\Big). \label{eq:Vlg2Vll}
\end{align} 
The variables $\theta^a_i,\theta^b_i$ represent the line-to-ground voltage angles for phase $a$ and $b$, respectively. The other two line-to-line voltage magnitudes $V_{i}^{bc}$ and $V_{i}^{ca}$ can be expressed using analogous expressions. Similar to the PVUR case, we define $z_{2,i}^L=0$ for non-three-phase buses. To minimize LVUR, we minimize the sum of $z_2^L$ across all buses, 
\begin{align}
   \min \sum_{i=1}^{n_b} z_{2,i}^L .  \label{eq:LVUR_obj}
\end{align}
LVUR constraints with upper bound $u_{LVUR}$ are expressed as 
\begin{align}
     z_2^L \leq u_{LVUR}. \label{eq:LVUR_const}
\end{align}

\subsection{TP-OPF Problem Formulations}
In this subsection, we describe the various TP-OPF problem formulations that represent typical optimization problems that a distribution system operator might be interested in solving. 

\subsubsection{Minimize losses}
Our first problem minimizes network losses without any consideration of voltage unbalance. The formulation for this problem can be summarized as
\begin{align}
    &\mathrm{min} &&\hspace{-4em}\mathrm{Network~Losses~\eqref{eq:loss}} \nonumber    \tag{P1-Loss}\\
    &\mathrm{s.t.} 
    &&\hspace{-4em}\mathrm{System~ Constraints ~\eqref{eq:Ypr}-\eqref{eq:pow_bal}} \nonumber
\end{align}

\subsubsection{Minimize voltage unbalance}
Our next problem formulations minimizes voltage unbalance according to the three voltage unbalance definitions. To minimize VUF, we solve
\begin{align}
    &\mathrm{min} &&\hspace{-4em}\mathrm{VUF}~\eqref{eq:VUF_obj} \nonumber    \tag{P2-VUF}\\
    &\mathrm{s.t.} 
    &&\hspace{-4em}\mathrm{System~ Constraints ~\eqref{eq:Ypr}-\eqref{eq:pow_bal}} \nonumber
\end{align}
The LVUR minimization problem is defined by
\begin{align}
    &\mathrm{min} &&\hspace{-4em}\mathrm{LVUR}~\eqref{eq:LVUR_obj} \nonumber    \tag{P3-LVUR}\\
    &\mathrm{s.t.} 
    &&\hspace{-4em}\mathrm{System~ Constraints ~\eqref{eq:Ypr}-\eqref{eq:pow_bal}} \nonumber\\
    &&&\hspace{-4em}\mathrm{Extra~constraints:}~\eqref{eq:LVUR_def1},~\eqref{eq:Vlg2Vll} \nonumber
\end{align}
while the PVUR minimization is solved with
\begin{align}
    &\mathrm{min} &&\hspace{-4em}\mathrm{PVUR}~\eqref{eq:PVUR_obj} \nonumber    \tag{P4-PVUR}\\
    &\mathrm{s.t.} 
    &&\hspace{-4em}\mathrm{System~ Constraints ~\eqref{eq:Ypr}-\eqref{eq:pow_bal}} \nonumber\\
    &&&\hspace{-4em}\mathrm{Extra~constraints:}~\eqref{eq:PVUR_z1},~\eqref{eq:PVUR_z2} \nonumber
\end{align}

\setcounter{figure}{1}
\begin{figure*}[b]
    \centering
    \includegraphics[width=1.00\textwidth]{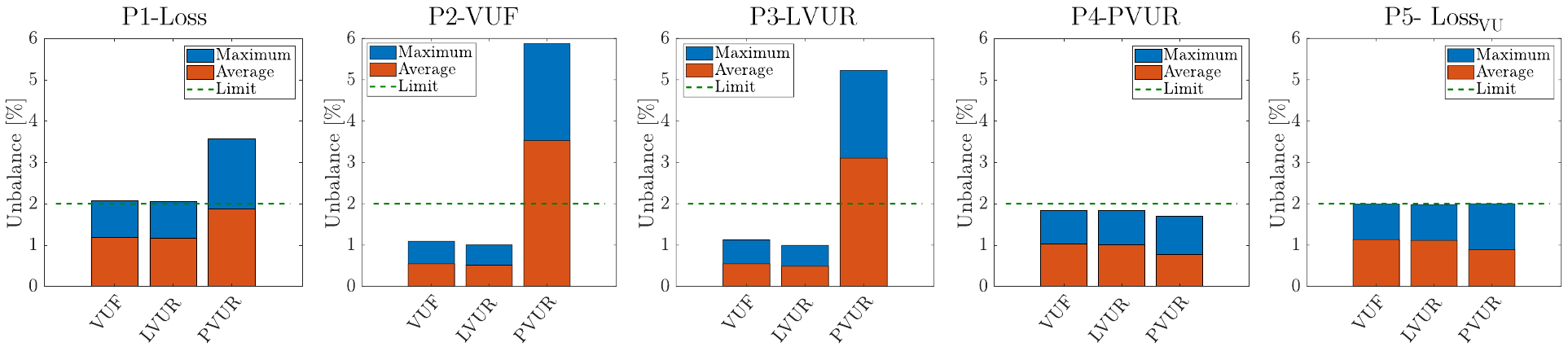}
    \caption{IEEE-13 bus feeder results. We show the values of the different voltage unbalance metrics for each of the considered optimization problem formulations.}
    \label{13VUlevel}
\end{figure*}

\subsubsection{Minimize losses with voltage unbalance constraints}
Our last problem formulation minimizes network losses with all the voltage unbalance definitions included as constraints, 
\begin{mini}|s|
    {}{\mathrm{Network~losses}~\eqref{eq:loss}}    
    {\tag{P5-Loss$_{VU}$}}{}
    \addConstraint{\mathrm{System~ Constraints:}~\eqref{eq:Ypr}-\eqref{eq:pow_bal}}
    \addConstraint{\mathrm{VUF~constraints:}~\eqref{eq:VUF_const}}
    \addConstraint{\mathrm{LVUR~constraints:}~\eqref{eq:LVUR_def1},~\eqref{eq:Vlg2Vll},~\eqref{eq:LVUR_const}}
    \addConstraint{\mathrm{PVUR~constraints:}~\eqref{eq:PVUR_z1},~\eqref{eq:PVUR_z2},~\eqref{eq:PVUR_const}},
\end{mini}
Using the various optimization models described previously, we analyze the impact of minimizing different objective functions on important network parameters, such as losses and voltage unbalance in the next section.

%% file: sections/Simulation.tex
\section{Simulation Results}
\label{sec:Results}

The three-phase AC optimal power flow problem is a large-scale, non-convex, non-linear optimization problem that is challenging to solve. We implemented the optimization problem in Julia \cite{julia} using JuMP \cite{DunningHuchetteLubin2017}, and used the solver Ipopt \cite{Ipopt}. 
We provide detailed analysis of both a small test case based on the IEEE 13-bus feeder \cite{schneider2017analytic}, as well as a larger test case based on one of the taxonomic distribution feeders from Pacific Northwest National Laboratory (PNNL) \cite{schneider2008modern}.

\subsection{IEEE 13-bus Feeder}
Fig.~\ref{13feeder} shows the modified IEEE 13-bus feeder with single-phase solar PV installations at seven nodes, illustrated by houses. The maximum apparent power rating of each single-phase solar PV inverter is 50 kVA. The PV penetration level, calculated as the ratio of total PV generation (in kW) to the total rated load (in kW), was chosen to be 35\%. For our analysis, we run the five optimization problem formulations: (P1-Loss), (P2-VUF), (P3-LVUR), (P4-PVUR) and (P5-Loss$_{VU}$). We compare the solutions in terms of voltage unbalance as defined by the three different metrics, in addition to the overall network losses and the power factor at the substation. In problem (P5-Loss$_{VU}$), we set the voltage unbalance limits to $u_{VUF}=u_{PVUR}=0.02$ and $u_{LVUR}=0.03$.

\setcounter{figure}{0}
\begin{figure}[t]
	\centering	        	
	\includegraphics[width=0.35\textwidth]{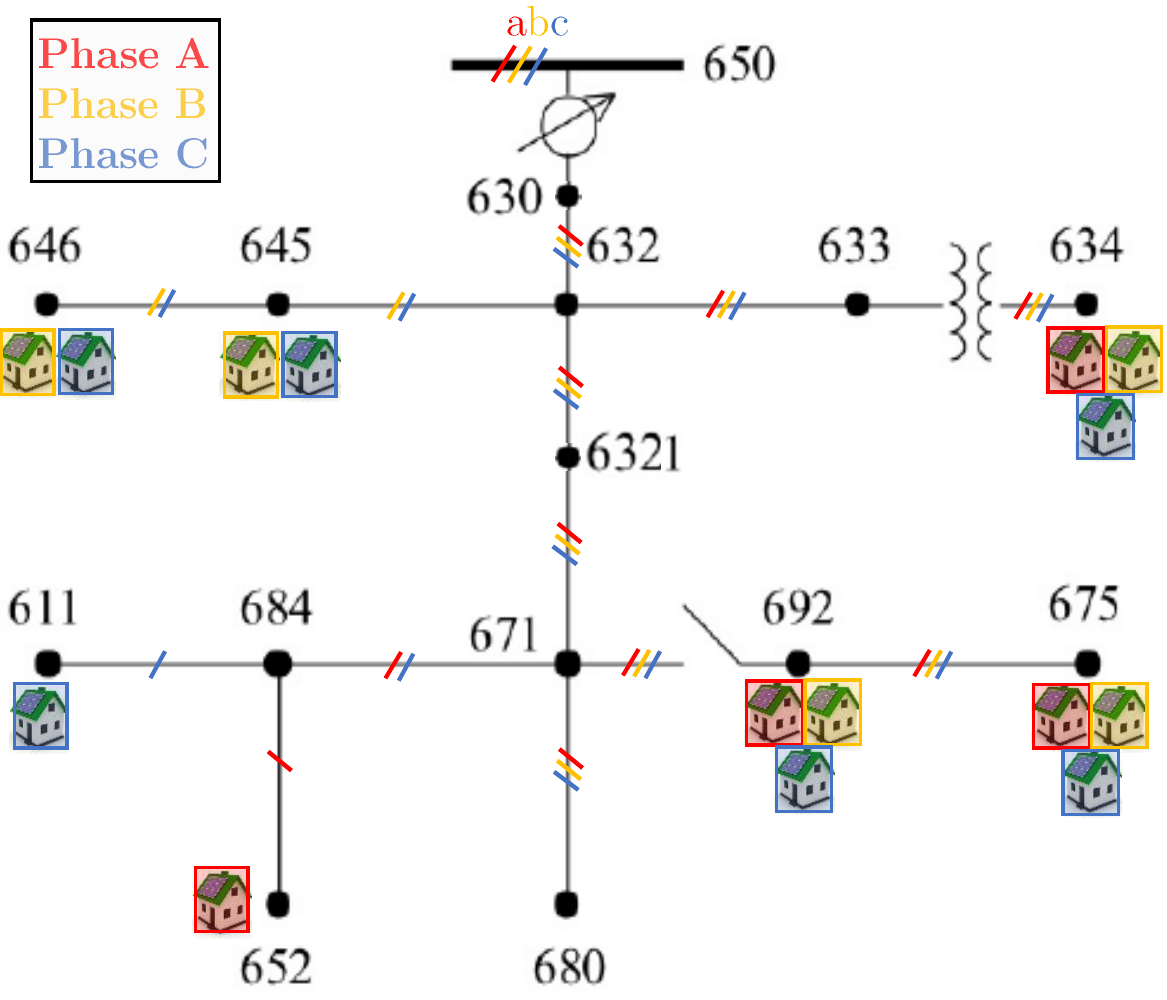}
	\caption{Modified IEEE-13 bus feeder}
	\label{13feeder}
\end{figure}

Fig.~\ref{13VUlevel} illustrates the average and maximum values of the various voltage unbalance metrics for each optimization problem solution. The maximum voltage unbalance level is calculated as the highest level among all the three-phase buses of the feeder. The average value is ratio of the sum of voltage unbalance at all three-phase buses to the number of three-phase buses. The green dashed line represents the allowable unbalance limit specified by IEC and IEEE standards for VUF and PVUR, respectively. The NEMA definition has a higher limit of 3\% which is never violated and hence also not shown in the figure. Other performance metrics to evaluate power quality such as network losses, substation power factor ($\cos\phi_{ss}$) are summarized in Table~\ref{13busSR} along with the average reactive power injections ($Q_{avg}^{inj}$). 

\setcounter{figure}{3}
\begin{figure*}[b]
    \centering
    \includegraphics[width=1.00\textwidth]{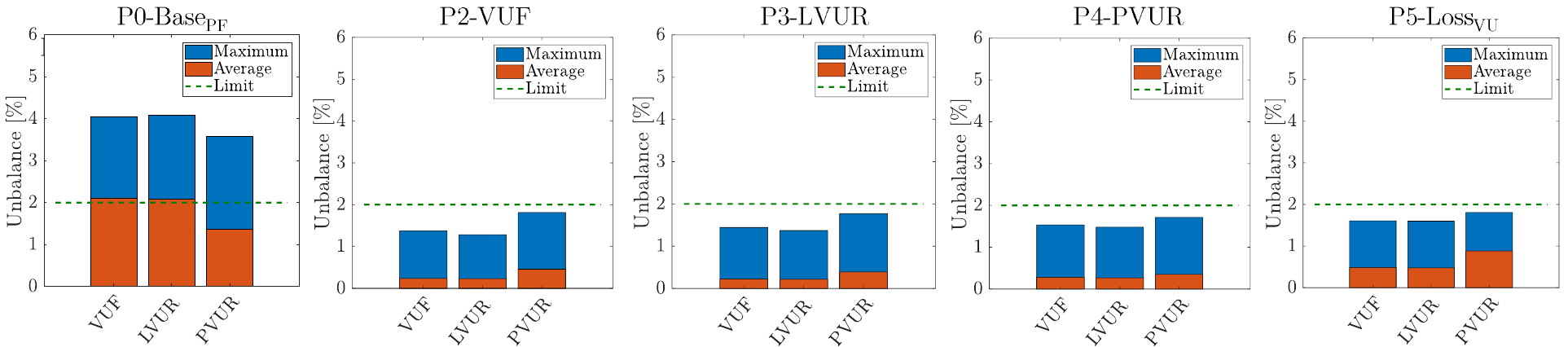}
    \caption{Taxonomic feeder: R2-12-47-2 results}
    \label{646VUlevel}
\end{figure*} 

Fig.~\ref{13VUlevel} shows that minimizing network losses leads to an operating condition where PVUR violates the unbalance limit. This highlights the need to consider voltage unbalance when we optimize system operations.
We further observe in Fig.~\ref{13VUlevel} and Table~\ref{13busSR} that VUF and LVUR are strongly related. When we minimize losses or any voltage unbalance metric, the resulting values for VUF and LVUR are similar. Another noteworthy observation is that PVUR in Fig.~\ref{13VUlevel}(b)-(c) significantly increases when we minimize VUF or LVUR. This shows that minimizing one voltage unbalance definition might lead to violation of another voltage unbalance definition. 

\begin{table}[t]
\renewcommand{\arraystretch}{1.3}
\centering
\caption{IEEE-13 bus feeder results for network losses, substation power factor and average reactive power injection}
\begin{tabular} {|c|c|c|c|}
\hline
\textbf{Problem} & \textbf{Loss} [kW] & \textbf{cos}$\phi_{ss}$  & \textbf{Q}$_{avg}^{inj}$ [kVAR] \\ \hline
P1-Loss &  88.63   & 0.96   & 41.41  \\ \hline    
P2-VUF &  114.67   & 0.86   & -16.08   \\ \hline 
P3-LVUR &  111.23   & 0.87   & -10.11  \\ \hline 
P4-PVUR &  92.48   & 0.93   & 20.75  \\ \hline 
P5-Loss$_{VU}$ &  91.01   & 0.93   & 23.76  \\ \hline 
\end{tabular}
\label{13busSR}
\end{table} 

By comparing the simulation results for problem (P1-Loss) with the results for problems (P2-VUF) and (P3-LVUR) in Table~\ref{13busSR}, we see that the average reactive power injection is negative when we minimize VUF or LVUR. This indicates that the solar PV inverters are absorbing reactive power from the system. It also explains the increase in  losses and reduction in power factor, since the substation needs to supply the additional reactive power and satisfy the power balance equation in~\eqref{eq:pow_bal}. In contrast, the average reactive power injection for (P4-PVUR) is positive, leading to higher power factor at the substation and lower network losses. 
Finally, we consider the results for problem (P5-Loss$_{VU}$), which minimizes losses subject to voltage unbalance constraints, in Table~\ref{13busSR}. We conclude that it is possible to obtain a solution which keeps the maximum values of all the unbalance metrics within the standards' limits, while incurring only a minimal increase in the losses.

\subsection{PNNL Taxonomic Feeder: R2-12-47-2}
The R2-12-47-2 feeder shown in Fig.~\ref{646feeder} comprises of 646 buses representing a moderately populated suburban area with mainly composed of single family homes and light commercial loads \cite{schneider2008modern}. The colored dots represent solar PV inverters connected to single-phase nodes in the feeder. Each inverter has a maximum rating of 10 KVA to achieve a PV penetration level of 50\% of the total rated load. 

\setcounter{figure}{2}
\begin{figure}[t]
	\centering	        	
	\includegraphics[width=0.4\textwidth]{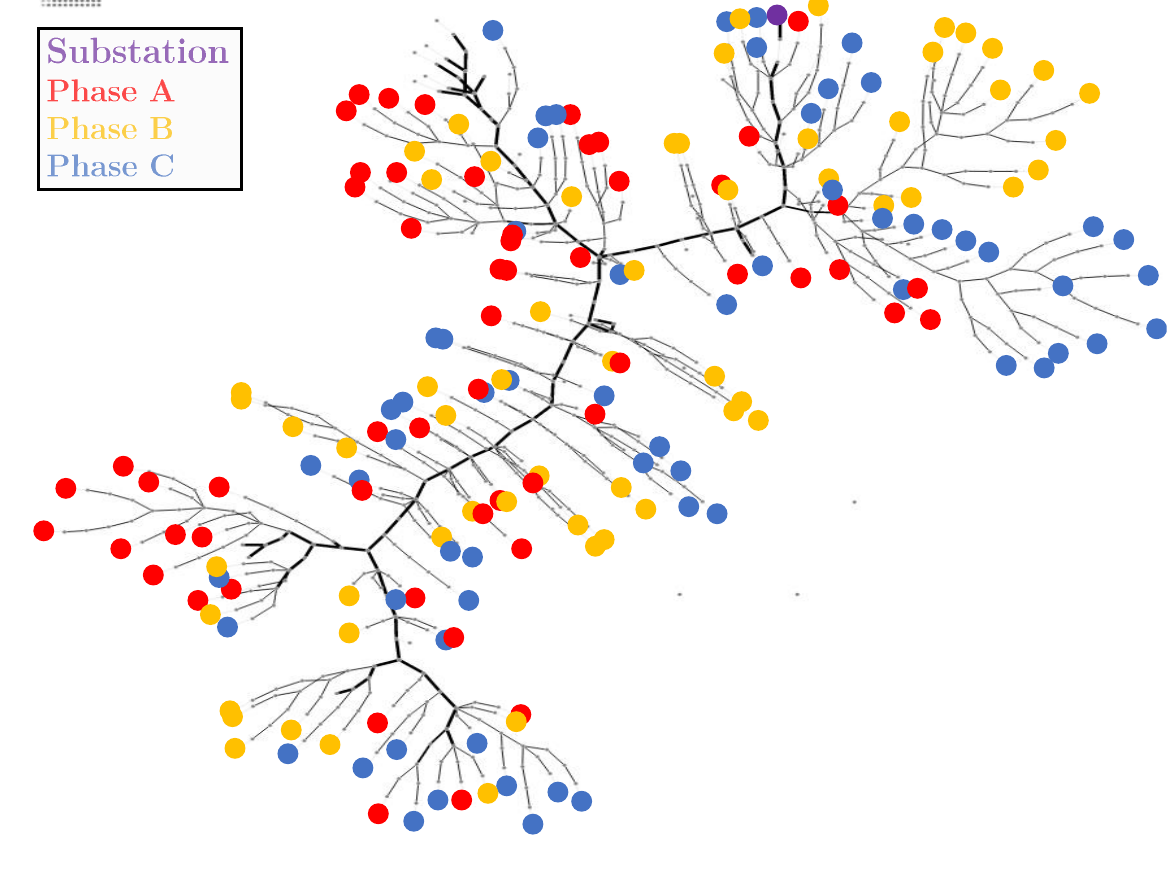}
	\caption{R2-12-47-2 taxonomic feeder}
	\label{646feeder}
\end{figure}

 Fig.~\ref{646VUlevel} illustrates the average voltage unbalance results for all cases and Table~\ref{646busSR} summarizes results for other power quality metrics as well as average reactive power injections. Different from the IEEE-13 bus results, the base case solution (P0-Base$_{PF}$) is not obtained by solving the TP-OPF optimization problem, but by solving the power flow in GridlabD using the Newton-Raphson method.


We again observe from the results in  Fig.~\ref{646VUlevel} and Table~\ref{646busSR} that minimizing any of the voltage unbalance metrics leads to similar results, with a reduction in voltage unbalance, losses and power factor relative to the base case. The average reactive power is also negative for all cases minimizing the voltage unbalance metrics, as shown in Table~\ref{646busSR}. 

Finally, we compare the results for problem (P5-Loss$_{VU}$) with other cases in  Fig.~\ref{646VUlevel} and Table~\ref{646busSR}. The average reactive power for (P5-Loss$_{VU}$) is positive, which results in significantly lower losses and higher power factor. The voltage unbalance metrics are well below their respective limits, indicating that minimizing losses is sufficient to maintain acceptable levels of voltage unbalance in this case.

\begin{table}[t]
\renewcommand{\arraystretch}{1.3}
\centering
\caption{R2-12-47-2 feeder results for network losses, substation power factor and average reactive power injection}
\begin{tabular} {|c|c|c|c|}
\hline
\textbf{Problem} & \textbf{Loss} [kW] & \textbf{cos}$\phi_{ss}$  & \textbf{Q}$_{avg}^{inj}$ [kVAR] \\ \hline
P0-Base$_{PF}$   &  367.32   & 0.86   & 0  \\ \hline    
P2-VUF &  84.90  & 0.82   & -1.31   \\ \hline 
P3-LVUR &  87.00   & 0.79   & -1.66  \\ \hline 
P4-PVUR &  90.90   & 0.75   & -2.15  \\ \hline 
P5-Loss$_{VU}$ &  30.30  & 0.98   & 0.81  \\ \hline 
\end{tabular}
\label{646busSR}
\end{table} 

%% file: sections/conclusion.tex
\section{Conclusion}
	\label{sec:Conclusion}
The goal of our paper is to evaluate the impact of minimizing different definitions of voltage unbalance when solving a three-phase AC optimal power flow (TP-OPF) problem. We present a full three-phase AC OPF formulation, which we extend with the different voltage unbalance metrics. 

Our case study demonstrates that care must be taken when minimizing voltage unbalance. In our results for the IEEE 13-bus test feeder, we observe minimizing VUF and LVUR lead to similar results, but we obtain a PVUR value beyond acceptable limits. The results obtained when minimizing losses subject to the voltage unbalance constraints showed that it is possible obtain a solution with low losses without violating any of the voltage unbalance limits. Our results on the larger taxonomic feeder are different. In this case, we compared our optimized solutions against a base case without reactive power injections from the PV inverters. We observe that minimizing \emph{any} of the voltage unbalance metrics leads to similar results, and a reduction in losses relative to the base case despite negative reactive power injections. In contrast, when we minimize losses (with voltage unbalance constraints) we obtain a solution with much lower losses and positive reactive power injections.

When solving three-phase optimal power flow, scalability of the optimization problems to large-scale, realistic cases is a challenge. Our ongoing and future work is focused on investigating numerically efficient approaches that can enable investigation of more realistic distribution systems.